\documentclass[
    ,final            
  ]
  {aipproc}

\usepackage{natbib}
\usepackage{amssymb}

\layoutstyle{8x11single}

\begin{document}
\newcommand{\sv}{$\sigma_{los}$} 
\newcommand{\rv}{$r_{200}$} 
\newcommand{\vv}{${\rm v}_{200}$} 
\newcommand{\mv}{$M_{200}$} 
\newcommand{\ks}{km~s$^{-1}$} 

\title{The evolution of mass profiles of galaxy clusters}


\author{A. Biviano}{
  address={INAF/Osservatorio Astronomico di Trieste, Italy}
}

\author{B. Poggianti}{
  address={INAF/Osservatorio Astronomico di Padova, Italy}
}

\begin{abstract}
  We determine the average mass profile of galaxy clusters at two
  different redshifts and compare its evolution with cosmological
  model predictions. We use two samples of galaxy clusters spanning
  similar (evolutionary corrected) mass ranges at different
  redshifts. The sample of low-redshift ($z \simeq 0.0-0.1$) clusters is
  extracted from the ESO Nearby Abell Cluster Survey (ENACS)
  catalog. The sample of high-redshift ($z \simeq 0.4-0.8$) clusters is
  mostly made of clusters from the ESO Distant Cluster Survey
  (EDisCS). We determine the average mass-profiles for these two
  cluster samples by solving the Jeans equation for hydrostatic
  equilibrium, using galaxies as tracers. By using two cluster galaxy
  populations, characterized by the presence and, respectively,
  absence of emission-lines in their spectra ('ELGs' and 'nELGs'
  hereafter), we are able to partially break the mass-profile
  orbital-anisotropy degeneracy.
  
  We find that the mass-profiles of both the nearby and the distant
  clusters are reasonably well fitted by a Navarro, Frenk \& White
  (NFW) model. The best-fit values of the NFW concentration parameter
  are as predicted by cosmological numerical simulations; cluster
  mass-density profiles become more concentrated with time. The
  evolution of the number-density profile of nELGs proceeds in the
  opposite sense, becoming less concentrated with time.

  In our analysis we also recover the orbital anisotropy of nELGs and
  ELGs . We find that in low-$z$ clusters nELGs follow almost
  isotropic orbits and ELGs have more radially-elongated orbits. In
  high-$z$ clusters both nELGs and ELGs follow radially-elongated
  orbits.

  We discuss these results in terms of the predicted secular mass
  growth of galaxy clusters and the transformation of ELGs into nELGs.
\end{abstract}

\maketitle

\section{Introduction}
\label{s:intro}
The study of the mass distribution within clusters and of its
evolution with redshift can provide very useful constraints on the
nature of dark matter (DM) \cite{SS00,Meneghetti+01,Reed+05} and on
the formation and evolution of galaxy clusters and their components
\cite{Springel+01,ElZant+04,Gao+04}. CDM cosmological simulations have
shown that the mass-density profiles of cosmological halos follow a
universal profile \cite{NFW97},
\begin{equation}
  \rho_{NFW} \propto (c \, r/r_{200})^{-1} (1 + c \, r/r_{200})^{-2},
\label{e:nfw}
\end{equation}
parameterized by the concentration parameter $c$
\footnote{The virial radius \rv~ is the radius
  within which the enclosed average mass density of a cluster is 200
  times the critical density.

  The virial mass \mv~ is the mass enclosed within a sphere of radius
  \rv.

The circular velocity is defined from the previous two quantities as
\vv=$(G$ \mv$/$\rv$)^{1/2}$.}. This universal model, named 'NFW' after
the initials of its proposers, is characterized by a central density
cusp.  The NFW model has been shown to fit reasonably well the
mass-density profile of nearby clusters
\cite[e.g.][]{BG03,Rines+03,KBM04,PAP05,BS06,Lokas+06}, although the
precise form of the profile near the center has been a matter of
debate \cite[e.g.][]{Moore+99,Hayashi+04,Diemand+05}.

Cosmological simulations predict a mild dependence of $c$ on the halo
mass, $M$, and a mild evolution of the $c-M$ relation with redshift,
$z$ \cite[e.g.][]{Gao+08,Duffy+08}.  Observed $c$ values of low-$z$
galaxy systems appear to be somewhat higher than theoretical
predictions, but the discrepancy mostly concerns low-mass galaxy
systems \cite{Duffy+08,Biviano08}.  Massive galaxy clusters at low-$z$
have the expected concentrations \cite[e.g.][]{KBM04,RD06}. This is
true also for clusters at redshift $z \approx 0.3$
\cite{vanderMarel+00}.  

Little is known about the mass-density profiles of galaxy clusters at
still higher-$z$. Mass-density profile concentrations have so far been
estimated for about a dozen clusters in total at $z \gtrsim 0.5$
\cite{ASF02,Hoekstra+02,Jee+05,VF06,Maughan+07,Rzepecki+07,SA07,Halkola+08},
all via gravitational lensing or deep X-ray observations. Here we
report on the determination of the average mass-density profile of 19
clusters at $0.39 \leq z \leq 0.8$ ($\overline{z}=0.56$), based on
the projected phase-space distribution of cluster galaxies, used as
tracers of the gravitational potential. 

We adopt $H_0=70$ km~s$^{-1}$~Mpc$^{-1}$, $\Omega_m=0.3$,
$\Omega_{\Lambda}=0.7$ throughout this paper.

\section{The samples} 
\label{s:data}
We use two samples of galaxy clusters, one at $\overline{z}=0.56$, the
EDisCS sample
\cite{Halliday+04,White+05,Poggianti+06,Desai+07,MilvangJensen+08},
and another at $\overline{z}=0.07$, the ENACS sample
\cite{Katgert+96,Katgert+98}, to investigate the evolution
of the average cluster mass profile.

We select the 15 EDisCS clusters with velocity dispersion \sv $\geq
250$ \ks, in order to have a more homogeneous data-set in terms of
mass. We then add to this sample four clusters from the MORPHS
\cite{Dressler+99,Poggianti+99}, all with masses in the same range
covered by the 15 EDisCS clusters. All 19 clusters have sufficiently
wide spatial coverage ($>0.5 \, r_{200}$) for the dynamical analysis,
as well as homogeneous photometry (which is needed for the
determination of the radial incompleteness, see below).

Among the ENACS clusters, we use the 59 ENACS clusters studied in
detail by \cite{Biviano+02,KBM04,BK04}.

We identify cluster members by the procedure described in
\cite{Biviano+06}, which has been validated on cluster-sized halos
extracted from cosmological numerical simulations
\cite{Biviano+06,Wojtak+07}. We then determine cluster line-of-sight
(los) velocity dispersions \sv~ by applying the robust biweight
estimator to the velocity distributions of selected cluster members
\cite{BFG90}. Cluster masses \mv~ are determined from the
\sv-estimates using the \sv--\mv~ relation of \cite{MM07}.

In summary, our data-set consist of 19 distant clusters from $z=0.39$
to $0.80$ ($\overline{z}=0.56$) with \mv~ masses from $0.7$ to $13.6
\times 10^{14} \, M_{\odot}$ (mean: $2.8 \times 10^{14} \, M_{\odot}$)
and 59 nearby clusters from $z=0.03$ to $0.10$ ($\overline{z}=0.07$)
with \mv~ masses from $0.4$ to $20.5 \times 10^{14} \, M_{\odot}$
(mean: $5.9 \times 10^{14} \, M_{\odot}$). When account is taken for
the predicted evolution in mass between $z=0.56$ and $z=0.07$
\cite{Adami+05,LC09} the two samples are found to contain halos of
similar (evolutionary-corrected) masses at two different cosmic epochs.

Determination of the average cluster mass profile requires stacking
together all the clusters in each of the two samples.  Stacking is
done by scaling the projected clustercentric galaxy distances, $R$, by
cluster virial radii, \rv, and the los galaxy velocities (in the
cluster rest frame), ${\rm v}_{rf} \equiv ({\rm v}-\overline{{\rm
    v}})/(1+\overline{{\rm v}}/c)$, by cluster circular velocities, \vv. In
such a way we avoid mixing up the virialized regions of the more
massive clusters with the unvirialized, external regions of the less
massive ones. Since cluster mass profiles are expected to depend very
mildly on cluster mass \cite[e.g.][]{NFW97,Dolag+04}, the shape of the
stacked cluster mass-profile is expected to be similar to the average
shape of the individual cluster mass profiles.

The center of each cluster is defined to be the position of its X-ray
surface-brightness peak, when available, or the position of its
brightest cluster galaxy otherwise.  We only use galaxies in the
radial range $0.05 \leq R/r_{200} \leq 1$ in the dynamical
analysis. At $R/r_{200}<0.05$ the analysis is unreliable because of
the uncertainty in the position of the cluster center, and at
$R>r_{200}$ the analysis is unreliable because dynamical relaxation is
not guaranteed outside the virial region.  In this radial range the
high-$z$ (low-$z$) stacked cluster contains 556 (respectively 2566)
galaxies.

We identify two populations of tracers: nELGs, the galaxies without
emission lines in their spectra, and ELGs, galaxies with emission
lines. More specifically, we classify ELGs the EDisCS galaxies with an
[OII] equivalent width $\geq 3$~\AA~ or with any other line in
emission \cite{Poggianti+06} and the MORPHS galaxies with a spectral
type different from 'k', 'k+a', and 'a+k' \cite{Poggianti+99}. We
refer to \cite{Katgert+96} for the ELG classification of ENACS
galaxies. The fraction of ELGs among the galaxies selected for the
dynamical analysis is 47\% in the high-$z$ sample, and only 13\% in
the low-$z$ sample, a difference that reflects the evolution of the
properties of cluster galaxies
\cite{Dressler+99,Poggianti+99,Poggianti+06}.

The samples are not spectroscopically complete to a given
magnitude. This is not a problem for the dynamical analysis as far as
the incompleteness does not depend on radius. The spectroscopic
incompleteness is indeed independent on radius for the ENACS
\cite{Katgert+98}.  On the other hand, for the high-$z$ cluster
sample we need to weigh galaxies according to their radial positions
when determining the galaxy number density profiles. The weighting
method is described in \cite{Poggianti+06}.

In constructing the galaxy number density profiles, we also need to
assign different weights to the contributions of different clusters to
the number counts in a given radial bin. This is because each cluster
contributes galaxies to the stacked sample only out to a limiting
radius, which is defined by the observational set-up. Hence, while at
small radii all clusters contribute, at large radii we need to correct
for those clusters that have not been sampled. The correction method
is similar to the one described in \cite{MK89} -- see
\cite{Biviano+02} and \cite{KBM04} for applications to the ENACS.

\begin{figure}
\includegraphics[totalheight=0.35\textwidth]{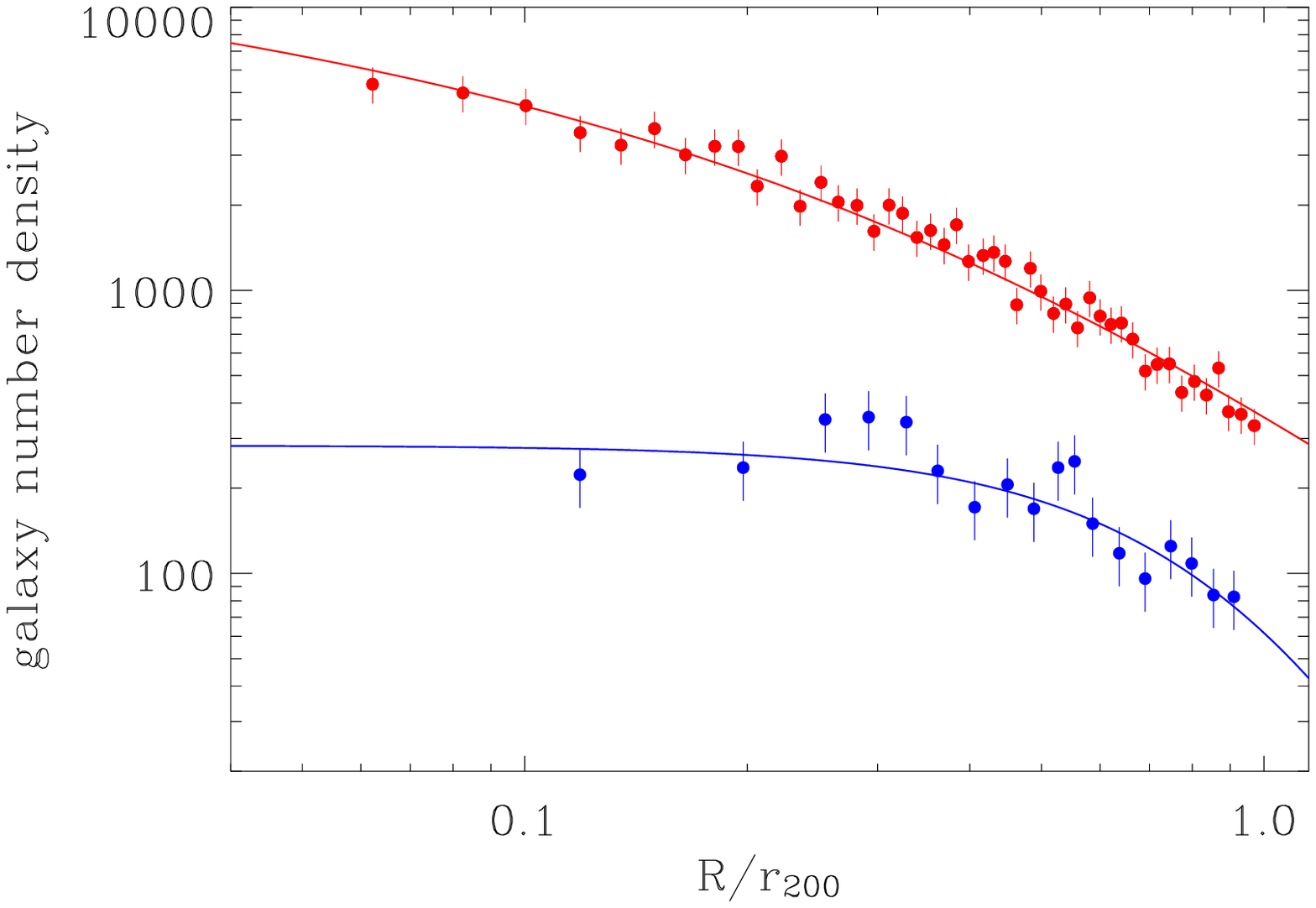}
\hfill
\includegraphics[totalheight=0.35\textwidth]{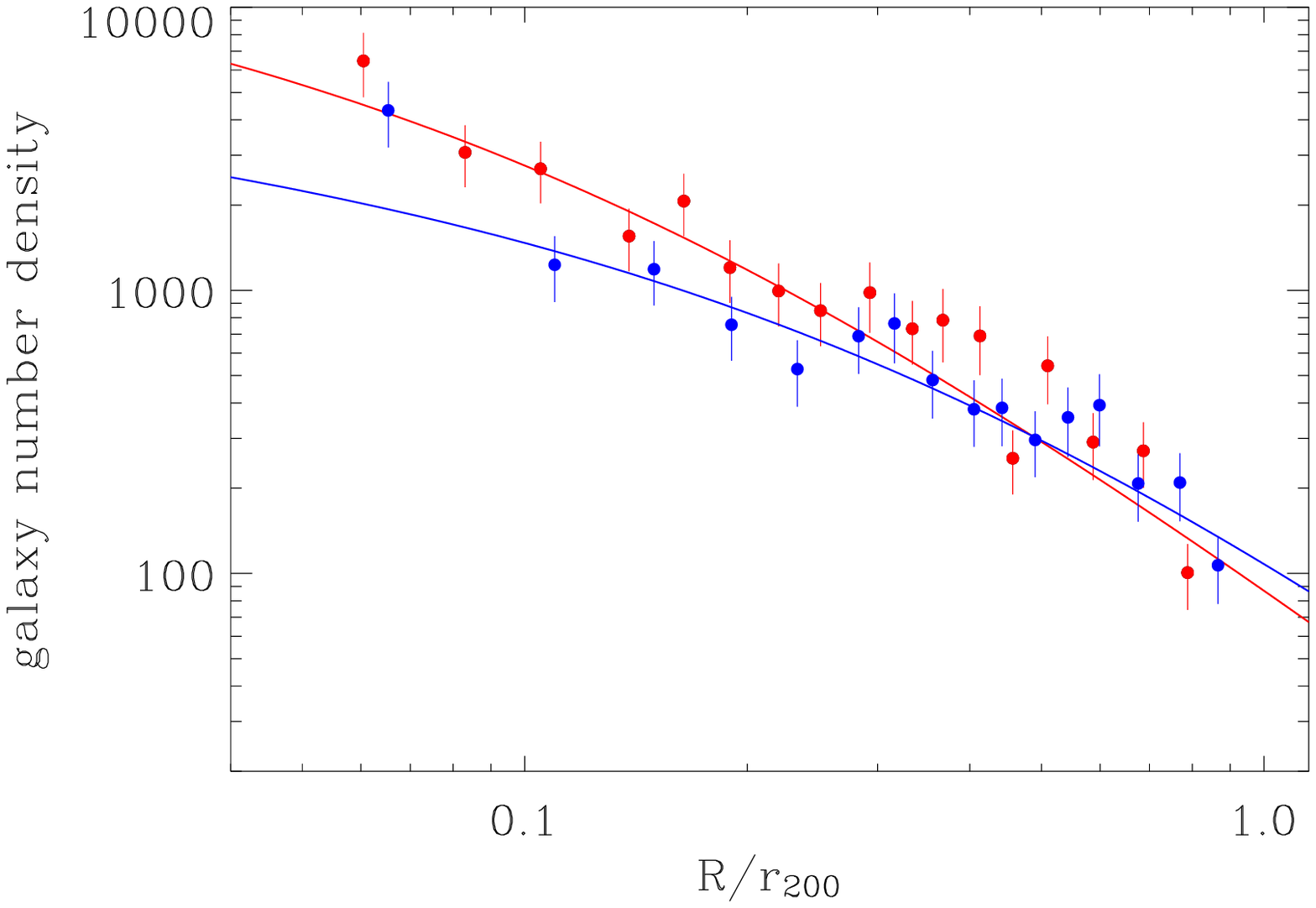}
\caption{Left panel: The projected number density profiles, $N(R)$,
  of the nELGs (red) and ELGs (blue) in the low-$z$ stacked
  cluster. Right panel: same as left panel, but for the high-$z$
  stacked cluster.  Solid lines represent best-fit models to the data,
  i.e. projected NFW models for nELGs and high-$z$ ELGs, and the core
  model for low-$z$ ELGs.}
\label{f:nprofs}
\end{figure}

\begin{figure}
\includegraphics[totalheight=0.35\textwidth]{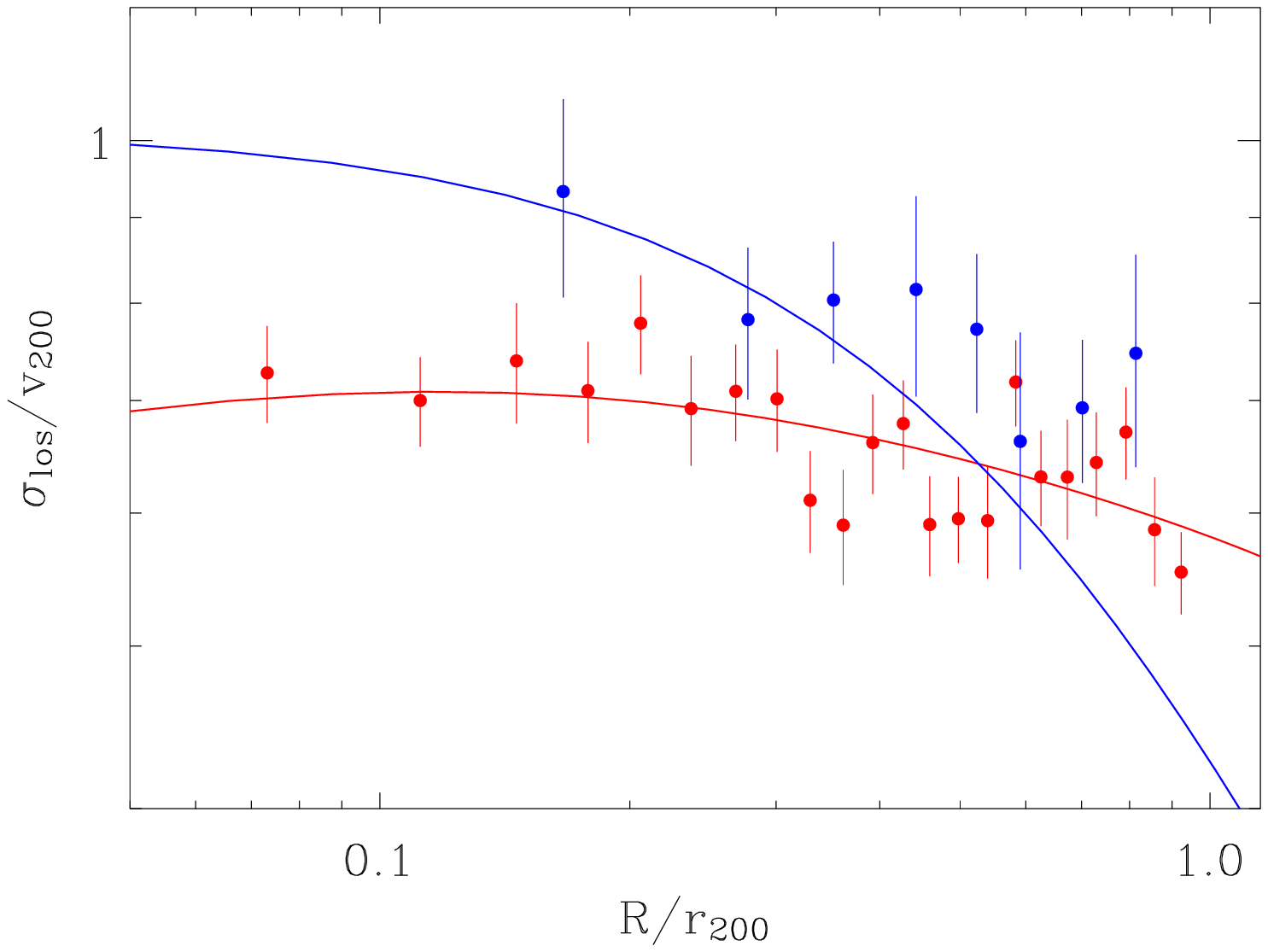}
\hfill
\includegraphics[totalheight=0.35\textwidth]{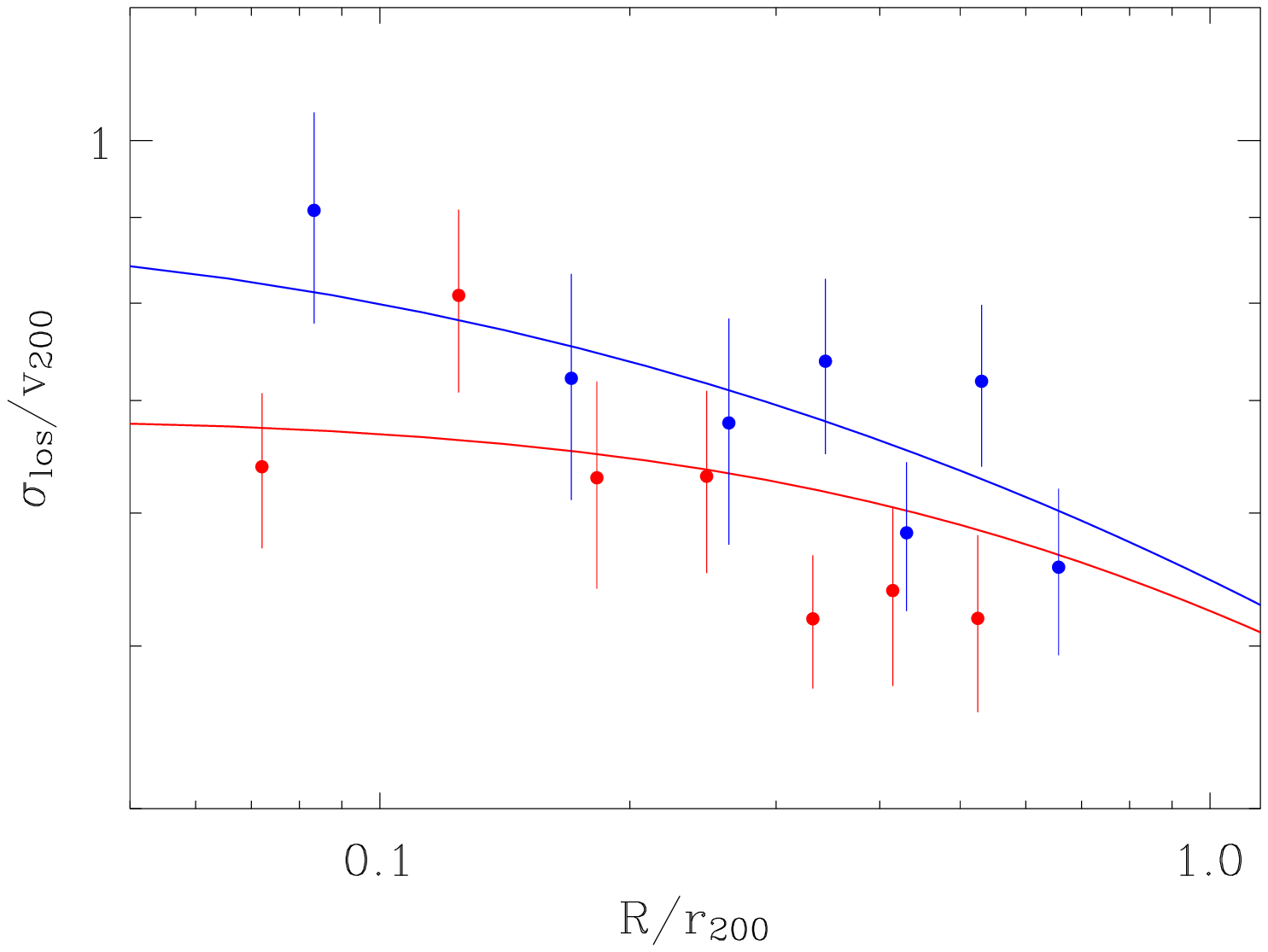}
\caption{Left panel: The los velocity dispersion profiles \sv$(R)$, of
  the nELGs (red) and ELGs (blue) in the low-$z$ stacked
  cluster. Right panel: same as left panel, but for the high-$z$
  stacked cluster.  Solid lines represent best-fit models to the data,
  i.e. a $c=4.0$ NFW mass model, $a/r_{200}=3.6$ (respectively
  $a/r_{200}=1.2$) OM velocity-anisotropy model for low-$z$ nELGs
  (respectively, ELGs), and a $c=3.2$ NFW mass model, $a/r_{200}=0.01$
  M{\L} velocity-anisotropy model for high-$z$ nELGs and ELGs.}
\label{f:vprofs}
\end{figure}

\section{The method}
\label{s:method}
The method we adopt for the dynamical analysis of our two stacked
clusters is based on the standard spherically-symmetric Jeans analysis
\cite{BT87}\footnote{Spherical symmetry is imposed by the stacking
  procedure, which is irrespective of the individual cluster position
  angles}.  In this analysis, the observables are the galaxy number
density profile $N(R)$ and the los velocity dispersion profile
$\sigma_{los}(R)$. These are displayed in Figures~\ref{f:nprofs} and
\ref{f:vprofs}, respectively, separately for the two cluster samples
and for the two cluster galaxy populations.

$N(R)$ is uniquely related to the 3-D galaxy number density profile
$\nu(r)$ via the Abel inversion equation.  The other observable,
$\sigma_{los}(R)$, is then uniquely determined if both the cluster mass
profile, $M(r)$, and the cluster velocity anisotropy profile,
$\beta(r)$, are known \cite{vanderMarel94,BM82}. The velocity
anisotropy profile $\beta(r)$ is
\begin{equation}
\beta(r) \equiv 1 - \frac{\rm{<} v_t^2 \rm{>}}{\rm{<} v_r^2 \rm{>}},
\label{eq:beta}
\end{equation}
where $\rm{<} v_t^2 \rm{>}$, $\rm{<} v_r^2 \rm{>}$ are the mean
squared tangential and radial velocity components, which reduce to
$\sigma_t^2$ and $\sigma_r^2$, respectively, in the absence of bulk
motions and net rotation (as we assume in the present analysis).

We adopt parameterized model representations of $M(r)$ and $\beta(r)$
and determine the best-fit parameters of these models by comparing the
observed $\sigma_{los}(R)$ profile with the predicted one, using the
$\chi^2$ statistics and the uncertainties on the observed profile. In
order to reduce the so-called ``mass--anisotropy'' degeneracy which
plagues these kinds of analyses \cite[see,
  e.g.,][]{Merritt87,vanderMarel+00,LM03} we adopt the method recently
suggested by \cite{Battaglia+08}. Namely, we consider two independent
tracers of the same gravitational potential, nELGs and ELGs, and
determine the best-fit parameters of $M(r)$ and $\beta(r)$ by a {\em
  joint} $\chi^2$-analysis of the best-fits to the $\sigma_{los}(R)$
profiles of the two populations. Clearly $M(r)$ must be the same for
both tracers, but $\beta(r)$ can in principle be different, so the
degeneracy is only partially broken; however the constraints on the
dynamics of the system are significantly stronger than when using a
single tracer.

Our choice of the $M(r)$ and $\beta(r)$ models is driven by the
results of the analysis of cluster-sized halos extracted from
cosmological numerical simulations. We adopt the NFW model,
eq.~(\ref{e:nfw}), parameterized by the concentration $c$, for the
mass-density profile and also for the galaxy number-density profile
\footnote{More precisely, we fit the {\em projected} NFW
  profile \cite{Bartelmann96} to $N(R)$.}, but we allow different
concentrations for the galaxy and the mass distributions.  Only when
the projected-NFW model does not provide an acceptable fit to
$N(R)$, we consider an alternative model,
\begin{equation}
N \propto [1+(R/R_c)^2]^{-\alpha}.
\label{e:core}
\end{equation}
We refer to this model\cite{CF78} as the 'core' model, since it is
characterized by a central constant density.

We consider two models for the velocity-anisotropy profile $\beta(r)$.
One is the Mamon-{\L}okas
('M{\L}' hereafter) model \cite{ML05b}
\begin{equation}
\beta = 0.5 \, r/(r+a), 
\label{e:ml}
\end{equation}
and the other is the Osipkov-Merritt 
('OM' hereafter) model \cite{Osipkov79,Merritt85-df}
\begin{equation}
\beta = r^2/(r^2+a^2), 
\label{e:om}
\end{equation}
Both the M{\L} and the OM models depend on just one free parameter,
the anisotropy radius, $a$, which marks the transition from the central
region, where $\beta \approx 0$ and the galaxy orbits are isotropic,
to the external region, where $\beta > 0$ and the galaxy orbits become
increasingly radial.

\section{Results}
\subsection{The nearby cluster sample}
\label{s:enacs}
The nELG $N(R)$ is best-fitted by a (projected) NFW profile with
$c=2.4$. A core model is required to fit the ELG $N(R)$ which avoid
the central cluster region, with best-fit parameter values
$R_c/r_{200}=1.28$ and $\alpha=3.2$. The best-fit models are displayed
in Figure~\ref{f:nprofs} (left panel). Abel-inversion of the $N(R)$
best-fitting models provides the 3-D number density profiles $\nu(r)$.

\begin{figure}
\includegraphics[totalheight=0.70\textwidth]{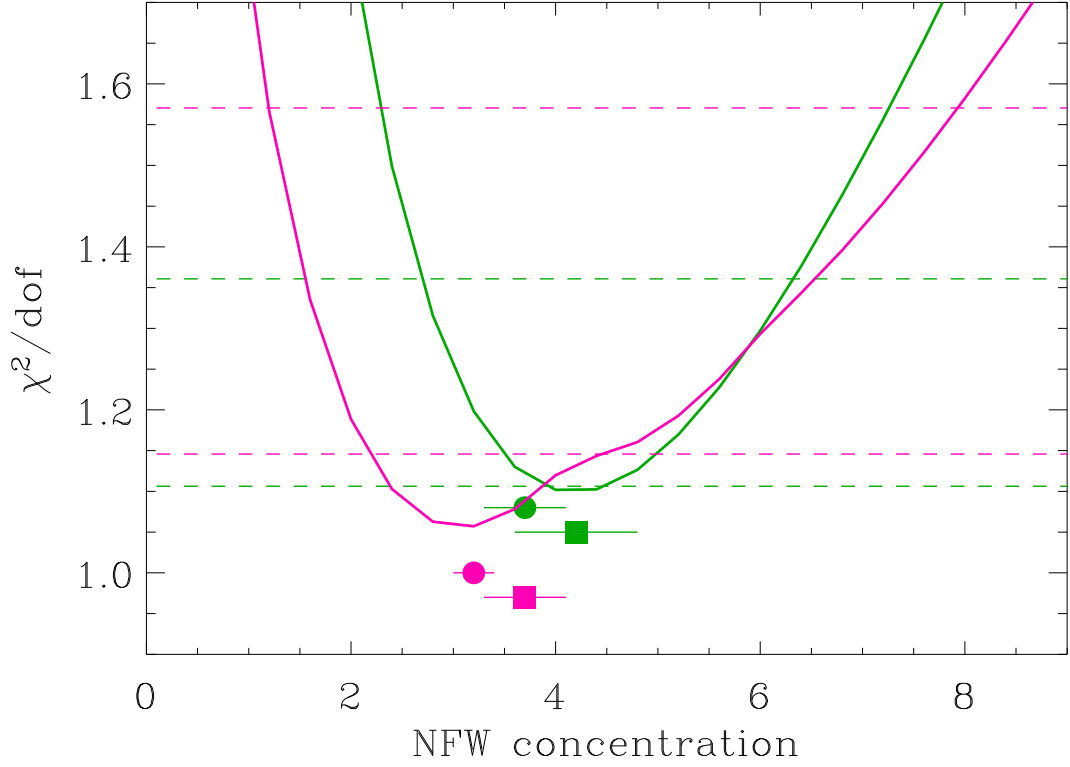}
\caption{The values of $\chi^2$ per degree of freedom obtained from
  the joint fit to the $\sigma_{los}(R)$ profiles of nELGs and ELGs for
  different values of the $c$ parameter of the NFW mass-profile model.
  The green and pink curves are for the low-$z$, and, respectively,
  the high-$z$ stacked cluster.  The two dashed lines indicate the 68\%
  and 90\% c.l. Symbols are theoretical predictions for the
  concentrations of the mass-density profiles of CDM halos with the
  same masses and redshifts as the clusters of our samples (squares:
  \cite{Gao+08}, dots: \cite{Duffy+08}; green: low-$z$, pink:
  high-$z$). The symbol error-bars are obtained by considering the
  1-$\sigma$ range in the distributions of cluster masses and
  average redshifts.}
\label{f:chi2c}
\end{figure}

The best-fit $c$ parameter of the stacked cluster NFW mass-density
profile is determined by a joint $\chi^2$ fit to the observed
\sv-profiles of the nELGs and ELGs.  The best-fit solution is obtained
when the OM model is adopted for the velocity anisotropy profiles of
the two galaxy populations. The best-fit value of the NFW
concentration parameter is $c=4.0_{-1.3}^{+2.3}$ (90\% confidence
levels, c.l. in the following), in agreement with \cite{KBM04}.  The
$\chi^2$ vs. $c$ solution is displayed in Figure~\ref{f:chi2c} (green
curve).

\begin{figure}
\includegraphics[totalheight=0.35\textwidth]{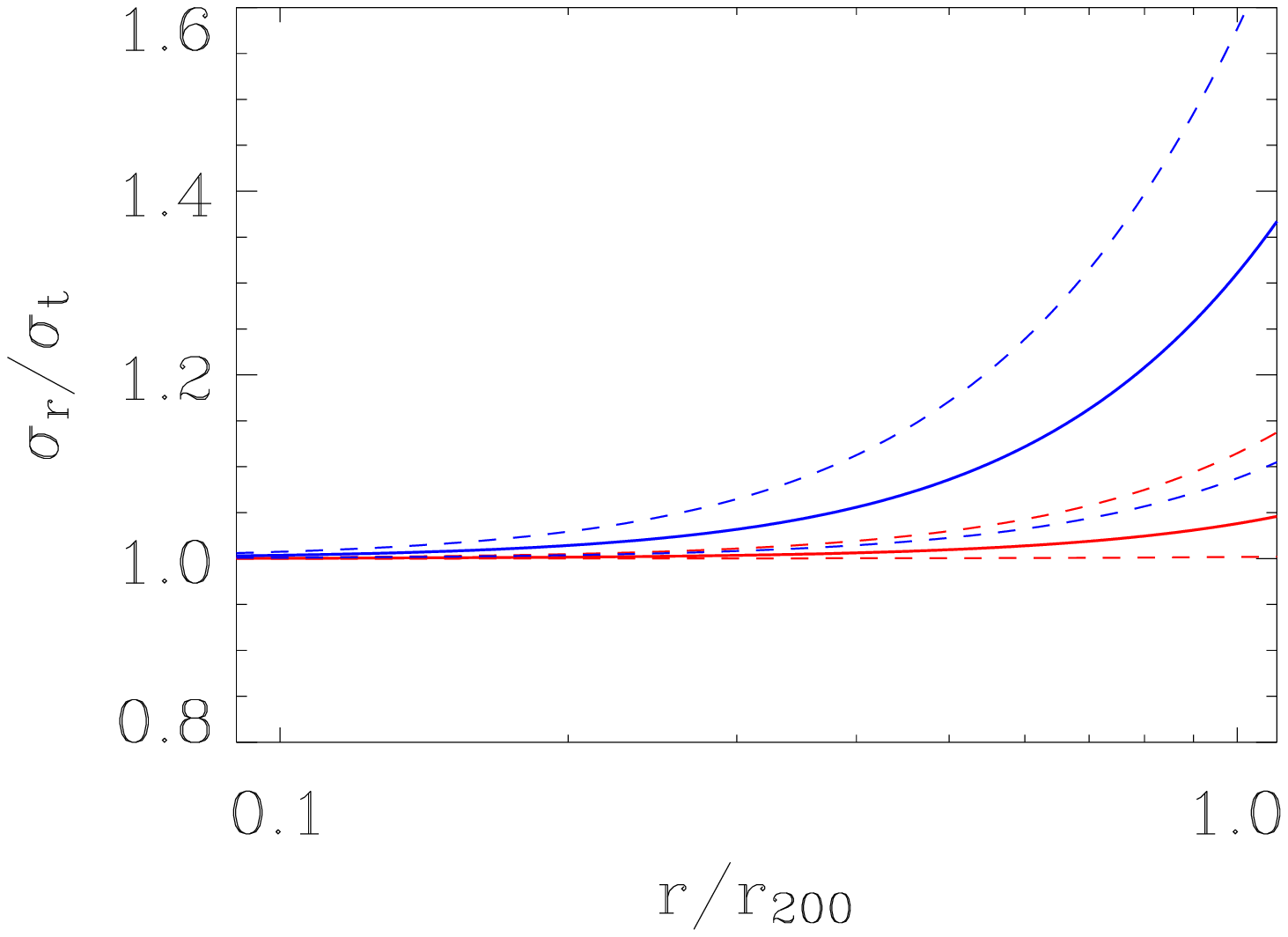}
\hfill
\includegraphics[totalheight=0.35\textwidth]{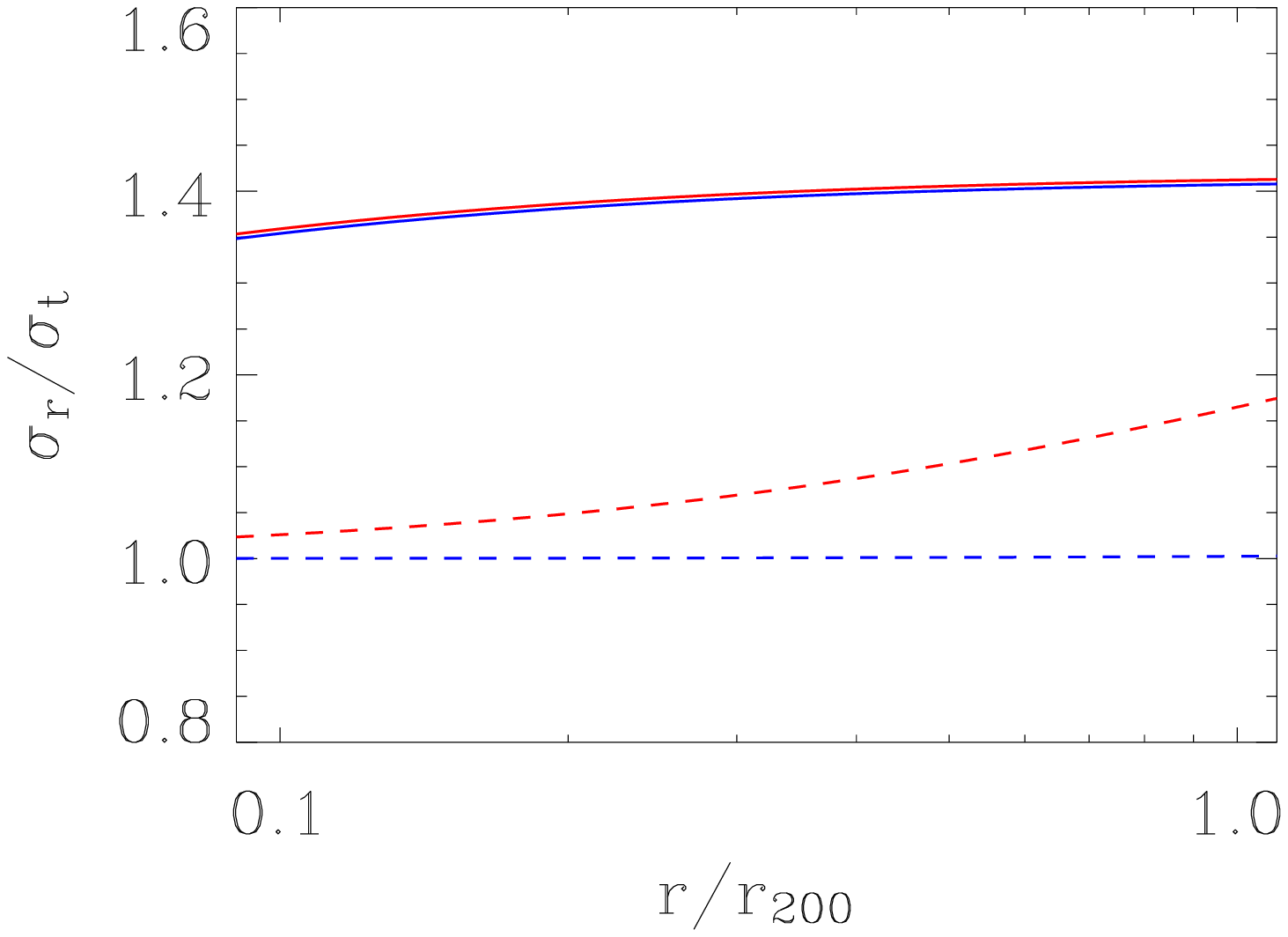}
\caption{Left panel: Best-fit velocity-anisotropy profile
  $\sigma_r/\sigma_t \equiv (1-\beta)^{-1/2}$ for nELGs (red curve)
  and ELGs (blue curve) in the low-$z$ stacked cluster. Dashed curves
  indicate 90\% c.l.  Right panel: same as left panel, but for the
  high-$z$ stacked cluster. In this case the best-fit solutions are at
  the lower-limit of the interval considered in the $\chi^2$
  minimization analysis, so only the lower c.l. to $\beta(r)$ are
  shown.}
\label{f:beta}
\end{figure}

Using this solution for the cluster mass profile we obtain the
best-fit $\beta(r)$ OM-model parameters $a/r_{200}=3.6_{-1.6}^{+13.4}$
and $a/r_{200}=1.2_{-0.4}^{+1.2}$ for the nELG and ELG populations,
respectively (90\% c.l.). These $\beta(r)$ profiles are shown in the
left panel of Figure~\ref{f:beta} [where we actually display
  $\sigma_r/\sigma_t \equiv (1-\beta)^{-1/2}$]. The best-fit
$\beta(r)$ solutions indicate that nELGs follow isotropic orbits
within the cluster virial region, and that ELG orbits are isotropic
near the center but become increasingly radial in the outer cluster
regions. These results are in agreement with \cite{BK04}.

In Figure~\ref{f:vprofs} (left panel) we display the observed los
velocity dispersion profiles, and the best-fit obtained via the Jeans
analysis.

\subsection{The distant cluster sample}
\label{s:ediscs}
Projected NFW models provide a good fit to the $N(R)$ of nELGs and
ELGs with best-fit parameters $c=7.5$ and $c=2.7$, respectively (see
the right panel of Figure~\ref{f:nprofs}).  Abel-inversion of these
best-fitting models then provide the 3-D number density profiles
$\nu(r)$ that we use in the dynamical analysis.

The joint best-fit to the los \sv-profiles of nELGs and ELGs is
obtained when the M{\L} $\beta(r)$ model is adopted. The mass profile
best-fit concentration value is $c=3.2_{-2.0}^{+4.6}$ (90\% c.l.). The
$\chi^2$ vs. $c$ solution is displayed in Figure~\ref{f:chi2c} (pink
curve). Using this solution for the stacked cluster mass profile, we
find that the best-fit value of the M{\L} $\beta(r)$ model parameter
$a$, is the same for the nELGs and for the ELGs, $a/r_{200}=0.01$, at
the lower limit of the $a$-range considered in the $\chi^2$
minimization analysis. The solution is only poorly constrained, $a<0.9
\, r_{200}$ and $a \leq 10.0 r_{200}$, for the nELGs and the ELGs,
respectively. The best-fit velocity-anisotropy profiles are shown in
Figure~\ref{f:beta} (right panel). Both cluster galaxy populations are
characterized by radially anisotropic orbits, although isotropy cannot
be formally excluded for the ELGs, given the uncertainties.

The observed \sv-profiles are shown in the right panel of
Figure~\ref{f:vprofs}, together with the best-fit solutions from the
Jeans analysis.

\section{Discussion and perspectives}
\label{s:disc}
Using cluster galaxies as tracers of the gravitational potential, we
have solved the Jeans analysis for the dynamical equilibrium of a
spherical system, and determined the average mass-density profiles of
galaxy clusters at $z \simeq 0.1$ and at $z \simeq 0.6$.  These
mass-density profiles are well described by NFW models with $c$ values
which decrease with increasing $z$. The best-fit $c$ values are in
agreement with the theoretical predictions of $\Lambda$CDM
models. This can be seen in Figure~\ref{f:chi2c} where we display our
best-fit results together with the theoretically predicted mean $c$
values.  These are determined by using the masses and redshifts of the
clusters in our two samples and applying the theoretical $c=c(M,z)$
relations of \cite{Gao+08} and \cite{Duffy+08}.

Our result appears to disagree with the conclusions of \cite{Duffy+08}
who claim that the observed $c$ values of clusters and groups are
significantly above the theoretically expected ones for group- and
cluster-sized cosmological halos. However most of the claimed
discrepancy is for low-mass galaxy systems, which are not covered by
the present analysis. Preliminary dynamical analyses of the
mass-density profiles of low-$z$ groups \cite{Biviano08}, done with
similar techniques as the one employed here, do indicate a discrepancy
in the direction reported by \cite{Duffy+08}.

An additional result of our dynamical analysis is the determination of
galaxy orbits in clusters at low-$z$ and high-$z$. We find that these
orbits become more isotropic with time. While low-$z$ cluster nELGs
have nearly isotropic orbits, high-$z$ cluster nELGs move on radially
elongated orbits, and so do both low-$z$ and high-$z$ ELGs (compare
the left-hand and right-hand panels of Figure~\ref{f:beta}).  

Orbital isotropization might result from the hierarchical accretion
process of clusters \cite{Gill+04} which undergo an initial, fast
accretion phase, followed by a slower, smoother accretion phase
\cite{LC09}.  During the fast accretion phase clusters are subject to
rapid variations of their gravitational potential
\cite{Manrique+03,PDdFP06,Valluri+07}, and these are capable of
isotropizing galaxy orbits
\cite{Henon64,LyndenBell67,KS03,Merritt05,LC09}. The end of the fast
accretion phase for cluster-sized halos occurs at $z \approx 0.4$
\cite{LC09}, hence it is not over yet for most of the clusters of our
high-$z$ sample. This can explain why the orbits of high-$z$ cluster
galaxies have not become isotropic yet.  At lower-$z$, galaxies that
have entered the cluster environment lately may experience only a
slower orbital isotropization process, probably caused by ram-pressure
\cite{Dolag+09}. Cluster ELGs are in this situation, while low-$z$
cluster nELGs have developed isotropic orbits already because they
were accreted much earlier than ELGs.

The different relative fraction of nELGs and ELGs in low- and high-$z$
clusters suggest that ELGs gradually transform into nELGs
\cite{Poggianti+06}. Since ELGs have a wider spatial distribution in
clusters than nELGs, as ELGs join the nELG population, the global nELG
spatial distribution become less concentrated. This is indeed observed
in our data-sets (compare the red curves in the left and right panels
of Figure~\ref{f:nprofs}).

The results presented here are still preliminary, as their statistical
significance is not very strong given the rather limited size of the
data-set for high-$z$ cluster galaxies. Substantial improvement in
the understanding of the accretion and internal dynamics history of
galaxy clusters requires much better statistics. In this sense, the
proposed ESA mission 
``EUCLID''\footnote{\texttt{http://sci.esa.int/science-e/www/area/index.cfm?fareaid=102}}
for the measurement of Dark Energy is extremely promising.
As of this writing, the EUCLID mission proposal considers a 20,000
deg$^2$ spectroscopic survey with a near-IR slitless spectrometer down
to a sensitivity level of $4 \times 10^{-16}$ erg cm$^{-2}$ s$^{-1}$
for emission-lines, with a 1/3 sampling rate. Such a survey should
return $\simeq 10,000$ clusters, each with $>20$ cluster members with
measured redshifts, at $0.5 \leq z \leq 0.8$. By stacking together
clusters of similar mass it should be possible to constrain not only
the evolution of the average cluster mass-density profile to an
accuracy of $\sim 3$\% in $c$, but also to measure the redshift
evolution of the $c=c(M)$ relation. With the same sample it will also
be possible to determine the evolution of the ELG orbits in clusters,
and to infer the history of cluster mass accretion.

\begin{theacknowledgments}
We acknowledge useful discussions with Giuseppina Battaglia, Alfonso
Cavaliere, Florence Durret, Marisa Girardi, and Gary Mamon. This
research has been financially supported from the National Institute
for Astrophysics through the PRIN-INAF scheme. This research has made
use of NASA's Astrophysics Data System.
\end{theacknowledgments}

\bibliographystyle{aipproc}   

\bibliography{biviano}

\IfFileExists{\jobname.bbl}{}
 {\typeout{}
  \typeout{******************************************}
  \typeout{** Please run "bibtex \jobname" to obtain}
  \typeout{** the bibliography and then re-run LaTeX}
  \typeout{** twice to fix the references!}
  \typeout{******************************************}
  \typeout{}
 }

\end{document}